\documentclass{iaus}
\usepackage{graphicx}
\usepackage{subfigure}

\title[Ices in starless and starforming cores] 
{Ices in starless and starforming cores}

\author[Karin I. \"Oberg et al.]   
{Karin I. \"Oberg$^1$,
A.~C. Adwin Boogert$^2$,
Klaus M. Pontoppidan$^3$,
Saskia van den Broek$^4$,
Ewine F. van Dishoeck$^4$,
Sandrine Bottinelli$^{4,5}$,
Geoffrey A. Blake$^6$
\and
Neal J. Evans II$^7$}

\affiliation{$^1$Harvard-Smithsonian Center for Astrophysics, 60 Garden St, Cambridge, MA 02139, USA\\
$^2$IPAC, NASA Herschel Science Center, Caltech, Pasadena, CA 91125, USA \\
$^3$Space Telescope Science Institute, 3700 San Martin Drive, Baltimore, MD 21218, USA\\
$^4$Leiden Observatory, Leiden University, PO Box 9513, 2300 RA Leiden, the Netherlands\\
$^5$CESR, CNRS-UMR 5187, 9 ave. Colonel Roche, BP 4346, 31028 Toulouse Cedex 4, France\\
$^6$Caltech, Division of Geological and Planetary Sciences, Pasadena, CA 91125, USA\\
$^7$Dep. of Astronomy, UT Austin, 1 University Station C1400, Austin, TX 78712, USA\\}

\pubyear{2011}
\volume{280}  
\jname{The Molecular Universe}
\editors{J. Cernicharo \& R. Bachiller, eds.}
\begin{document}

\maketitle

\begin{abstract}
Icy grain mantles are commonly observed through infrared spectroscopy toward dense clouds, cloud cores, protostellar envelopes and protoplanetary disks. Up to 80\% of the available oxygen, carbon and nitrogen are found in such ices; the most common ice constituents -- H$_2$O, CO$_2$ and CO -- are second in abundance only to H$_2$ in many star forming regions. In addition to being a molecular reservoir, ice chemistry is responsible for much of the chemical evolution from H$_2$O to complex, prebiotic molecules. Combining the exisiting ISO, Spitzer, VLT and Keck ice data results in a large sample of ice sources ($\sim$80) that span all stages of star formation and a large range of protostellar luminosities ($<$0.1--10$^5$ L$_\odot$). Here we summarize the different techniques that have been applied to mine this ice data set on information on typical ice compositions in different environments and what this implies about how ices form and evolve during star and planet formation. The focus is on how to maximize the use of empirical constraints from ice observations, followed by the application of information from experiments and models. This strategy is used to identify ice bands and to constrain which ices form early during cloud formation, which form later in the prestellar core and which require protostellar heat and/or UV radiation to form. The utility of statistical tests, survival analysis and ice maps is highlighted; the latter directly reveals that the prestellar ice formation takes place in two phases, associated with H$_2$O and CO ice formation, respectively, and that most protostellar ice variation can be explained by differences in the prestellar CO ice formation stage. Finally, special attention is paid to the difficulty of observing complex ices directly and how gas observations, experiments and models help in constraining this ice chemistry stage.
\end{abstract}

\firstsection 
\section{Introduction}

\noindent The evolution from starless to star-forming cores and further on to (extra)solar systems is accompanied by a rich chemistry, which will affect planet and planetesimal compositions. Much of this chemical evolution takes place on interstellar grain surfaces, in icy mantles. Ices are generally observed in the cold and dense interstellar medium that are the nurseries of stars and are also abundant in our own solar system in comets and other outer solar system bodies. Such icy bodies are proposed to have delivered volatiles to the young earth. Understanding their link to the chemistry observed during star formation is key to constrain the amount of organic material delivered to Earth as well as to exo-planets.

The first ices were detected in the interstellar medium almost 40 years ago by \cite{Gillet73}.  H$_2$O and CO were established early on as common ice constituents with abundances reaching 10$^{-4}$ $n_{\rm H_2}$. This makes ices the most common molecules after H$_2$ in many star forming regions. CO$_2$ is the third major ice constituent. Because of abundant atmospheric CO$_2$, it was only detected in the ISM after the launch of IRAS by \cite{dHendecourt89}. Following IRAS, the {\it Infrared Space Observatory} (ISO) and the {\it Spitzer Space Telescope} have, together with ground-based efforts, continued the ice exploration and interstellar ices. In addition to the major ice constituents, H$_2$O, CO and CO$_2$  are observed to often contain smaller amounts of CH$_3$OH, CH$_4$, NH$_3$ and OCN$^{-}$/XCN  (see \cite{Boogert04} for a review).  

The simple ices observed in the ISM may become the source of complex molecules as described by \cite{Charnley92} and \cite{Garrod08} and determining ice abundances and production channels in star forming regions is of great  interest for studies of prebiotic chemistry. Ices can form from direct freeze-out of molecules from the gas phase, through atomic addition on grain and ice surfaces, and through energetic processing of already existing ices. At the low temperatures ($<$20~K) prevalent in cloud cores, atoms and molecules will stick to the grain upon impact. In this stage grains are well protected from external UV radiation and atomic addition reactions probably dominate. \cite{Tielens82} worked out in detail how the atoms and molecules can be hydrogenated and oxygenated to form all commonly observed ices, except for CO, which condenses directly from the gas-phase. In this scheme, H$_2$O forms from O+H, CH$_3$OH from CO+H etc. CH$_3$OH can also form from UV processing of H$_2$O:CH$_4$ ice mixtures, however, and there has been some debate on which formation pathway dominates. \cite{Gibb04} and others have also suggested that at least some CO$_2$ and XCN form from thermal or UV processing, mainly based on high-mass protostellar studies with {\it ISO}.

{\it Spitzer} and complimentary ground-based surveys have more recently allowed us to observe ices toward large numbers of low-mass protostars and cloud cores, which have provided new observational constraints on how ice formation and evolution depend on the local environment. The aim of this contribution is to review the constraints supplied by ice observations toward both {\it ISO} and {\it Spitzer} targets, on the identifications of ice bands and of the main formation paths of the identified carriers. The figures are taken from \"Oberg et al. (2011, subm. to ApJ) unless otherwise noted.

\section{Ice observations}

\begin{figure}[b]
\begin{center}
 \includegraphics[width=4in]{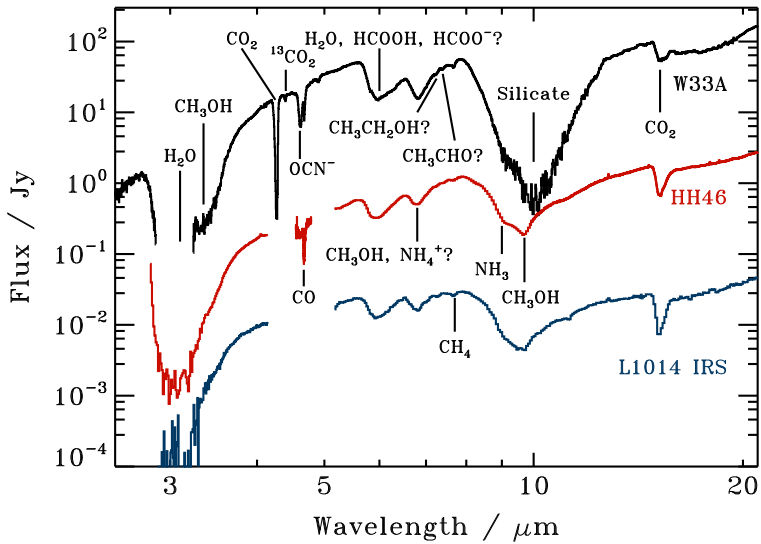} 
 \caption{Ice spectra toward the protostars W33A (10$^5$ L$_{\odot}$), HH46 (12 L$_{\odot}$) and L1014 IRS (0.09 L$_{\odot}$) from \cite{Gibb00,Boogert04b} and \cite{Boogert08}. The 3~$\mu$m portions of the spectra have been binned to increase the S/N.}
   \label{fig1}
\end{center}
\end{figure}

Ices are generally observed in absorption at mid-IR wavelengths and can therefore only be observed in lines of sight with infrared background sources. Protostars are excellent IR sources and ices in protostellar envelopes are the most extensively studied (Fig. 1). Early ice observations were exclusively ground based and thus limited to observe ice bands that coincide with atmospheric windows. IRAS and especially {\it ISO} opened up the full infrared spectral region, resulting in the first complete ice inventories toward bright protostars and a few cloud lines of sight. The {\it ISO} ice results are described in \cite{Gibb04} and these still constitute the largest high-mass protostellar sample.

Compared to previous space missions, \textit{Spitzer}'s high sensitivity enabled observations of ices toward low-mass protostars and also toward previously inaccessible background stars. The largest sample was obtained within the $c2d$ program, defined in \cite{Evans03}, which included IRS spectra of 50 low-mass protostars. The different ice bands were analyzed in a series of papers by
\cite{Boogert08}, \cite{Pontoppidan08}, \cite{Oberg08} and \cite{Bottinelli10}. 
Additional {\it Spitzer} observations exist on ices toward
background sources, looking through molecular clouds at a range of
extinctions (reported by \cite{Bergin05}, \cite{Knez05}, \cite{Pontoppidan06}, \cite{Whittet09} and \cite{Boogert11}) and toward protostars in the high-UV environment of IC 1396A by \cite{Reach09}. Because the  {\it Spitzer} spectrometer had a lower cutoff at $\sim$5~$\mu$m, complete ice inventories can only be obtained by adding complementary ground-based spectroscopy to cover the strong 3 $\mu$m H$_2$O, the 4.65
$\mu$m CO, the 4.60-4.62 $\mu$m XCN, and the 3.53 $\mu$m CH$_3$OH features and this was done using VLT and Keck. Results are summarized in \"Oberg et al. (2011, subm. to ApJ).

The combined {\it Spitzer}, ISO, VLT and Keck ice sample spans the full range of observed luminosities from the low-luminosity object L1014 IRS ($<$0.1L$_\odot$) to the $10^5$~L$_\odot$ massive protostar W33A (Fig. \ref{fig1}). The low-mass young stellar objects (YSOs) are located in  Perseus, Taurus, Serpens, Ophiuchus and Corona Australis, as well as a number of nearby isolated dense cores,
and thus cover star-forming regions with different histories and physical environments. The sample also represents a wide range of cloud, cloud core and YSO evolutionary stages, from envelope-dominated Class 0 sources to disks. 

\section{Ice identifications}

Figure \ref{fig1} shows ice spectra toward  protostars with the identifications of different bands to
H$_2$O, CO, CO$_2$, CH$_3$OH, NH$_3$, CH$_4$ and OCN$^-$ ice
marked. \cite{vanBroekhuizen05} found that the observed XCN band at 4.62 $\mu$m consists of two components. Based on laboratory spectroscopy from \cite{vanBroekhuizen04}, at least one of the components is due to OCN$^-$. OCN$^-$ in a separate ice matrix is probably the carrier of the second XCN band component (centered at 2175 cm$^{-1}$), but other identifications have also been suggested by e.g. \cite{Fraser05}.

\begin{figure}[htb]
\begin{center}
        \includegraphics[width=2.5in]{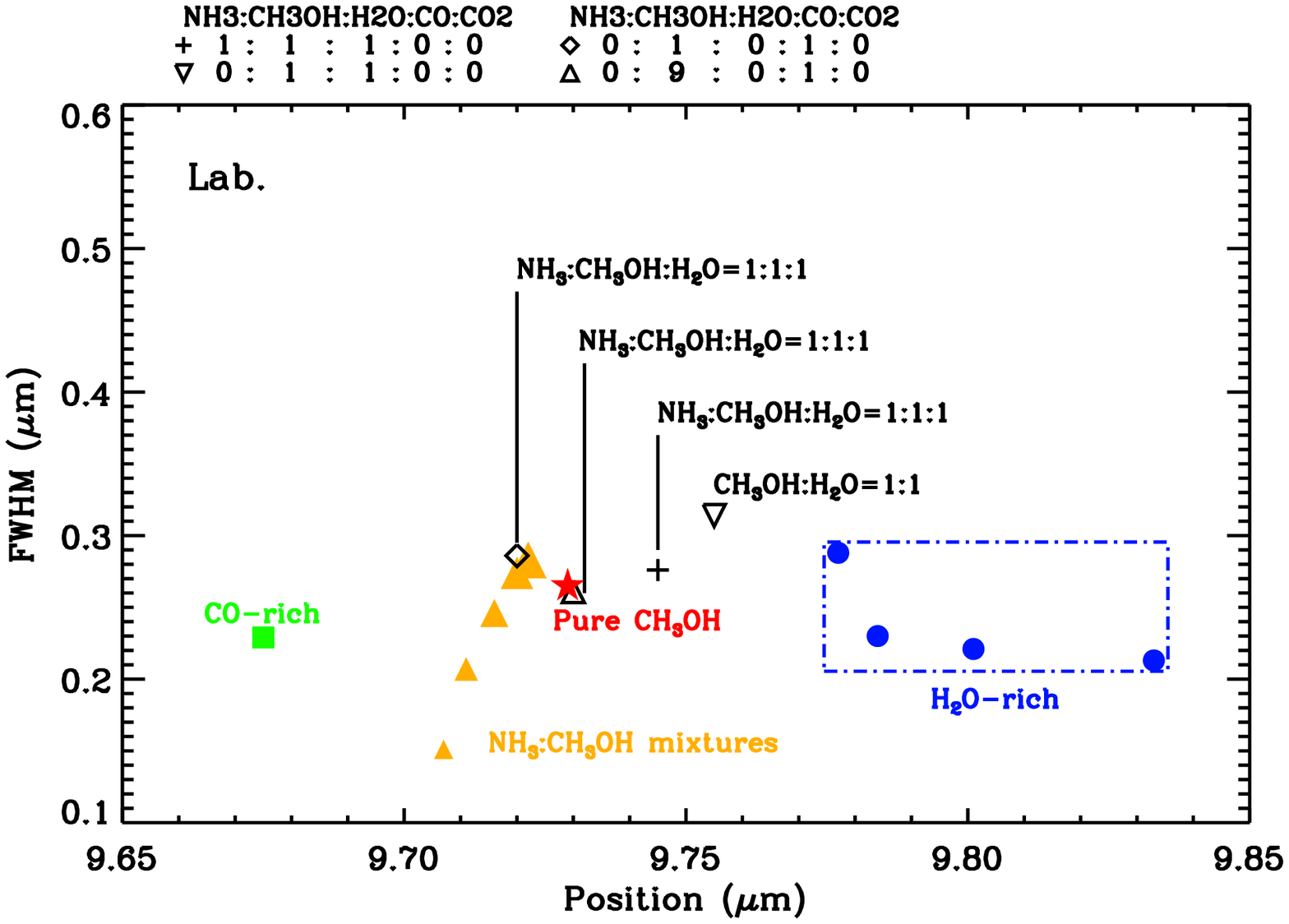}
        \includegraphics[width=2.5in]{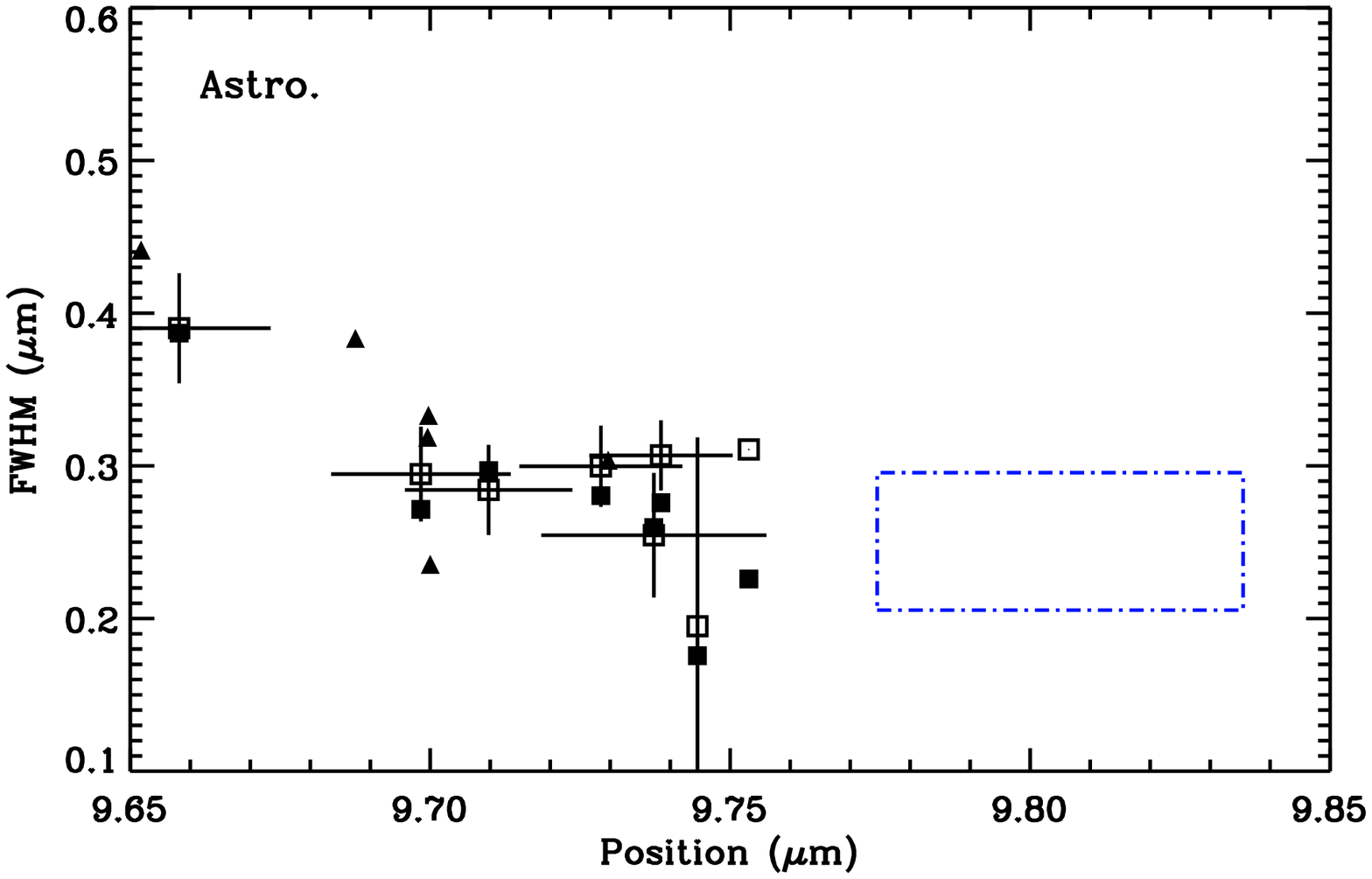}
    \caption{Figure from \cite{Bottinelli10} showing the FWHM and peak position of the CH$_3$OH feature in laboratory mixtures at 15 K (left panel) and in Spitzer $c2d$ spectra (right panel). In the right panel an increasing symbol size is indicative of increasing CH$_3$OH content. In both panels, the dash-dot polygons delimitate the parameter space of FWHM and peak positions corresponding to H$_2$O-rich mixtures. \label{fig2}}
    \end{center}
\end{figure}

In general the frequency of a solid state molecular vibration depends on the bonding environment of the vibrating molecule. This results in that the positions and shapes of ice spectroscopic features depend on the ice mixture(s).  Fig. \ref{fig2} from \cite{Bottinelli10} demonstrates how the  CH$_3$OH FWHM and band center depend on the ice mixture in laboratory experiments and how this can be used to constrain that protostellar CH$_3$OH ices are not present in H$_2$O-dominated ices.

\begin{figure}[htb]
\begin{center}
 \includegraphics[width=2in,angle=90]{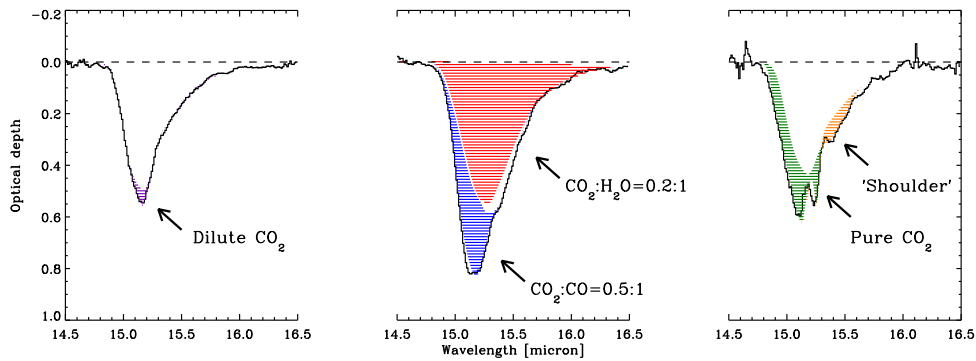} 
 \caption{Sketch of the five different components used to fit the CO$_2$ band on top of IRS 51, SVS 4-5, and RNO 91 spectra from \cite{Pontoppidan08}}
   \label{fig3}
\end{center}
\end{figure}

Some ice bands, such as CO and CO$_2$, cannot be fitted by a single laboratory ice mixture, but seem to trace molecules in two or more ice phases, often referred to as polar and apolar. Traditionally the observed spectra have been directly compared to a superposition of laboratory pure and mixed ice spectra to constrain ice components, see e.g. \cite{Merrill76}, \cite{Gibb04} and \cite{Zasowski09}. The constraints are often degenerate, however, since ice spectral features vary with ice composition, temperature and radiation processing. To address this, observed ice bands are now commonly decomposed phenomenologically and then the components are compared to laboratory spectra, see \cite{Tielens91}, \cite{Pendleton99}, \cite{Keane01} and \cite{Boogert08}. Because all observed spectra are decomposed into the same, small number of components, this method provides information on the sample as a whole, i.e. it directly shows which parts of the spectral profile are ubiquitous and which are environment dependent. Figure \ref{fig3} from \cite{Pontoppidan08} demonstrates how the derived CO$_2$ ice components can be identified with pure CO$_2$ ice, CO$_2$ mixed with CO (CO$_2$:CO), CO$_2$ mixed with H$_2$O ice
(CO$_2$:H$_2$O) and a shoulder associated with CO$_2$
mixed with CH$_3$OH.  \cite{Pontoppidan03} showed that the CO ice band can be similarly decomposed into three components corresponding to pure CO ice, CO:CO$_2$ and CO:H$_2$O or CO:CH$_3$OH.

\begin{figure}[htp]
\begin{center}
 \includegraphics[width=2.7in,angle=90]{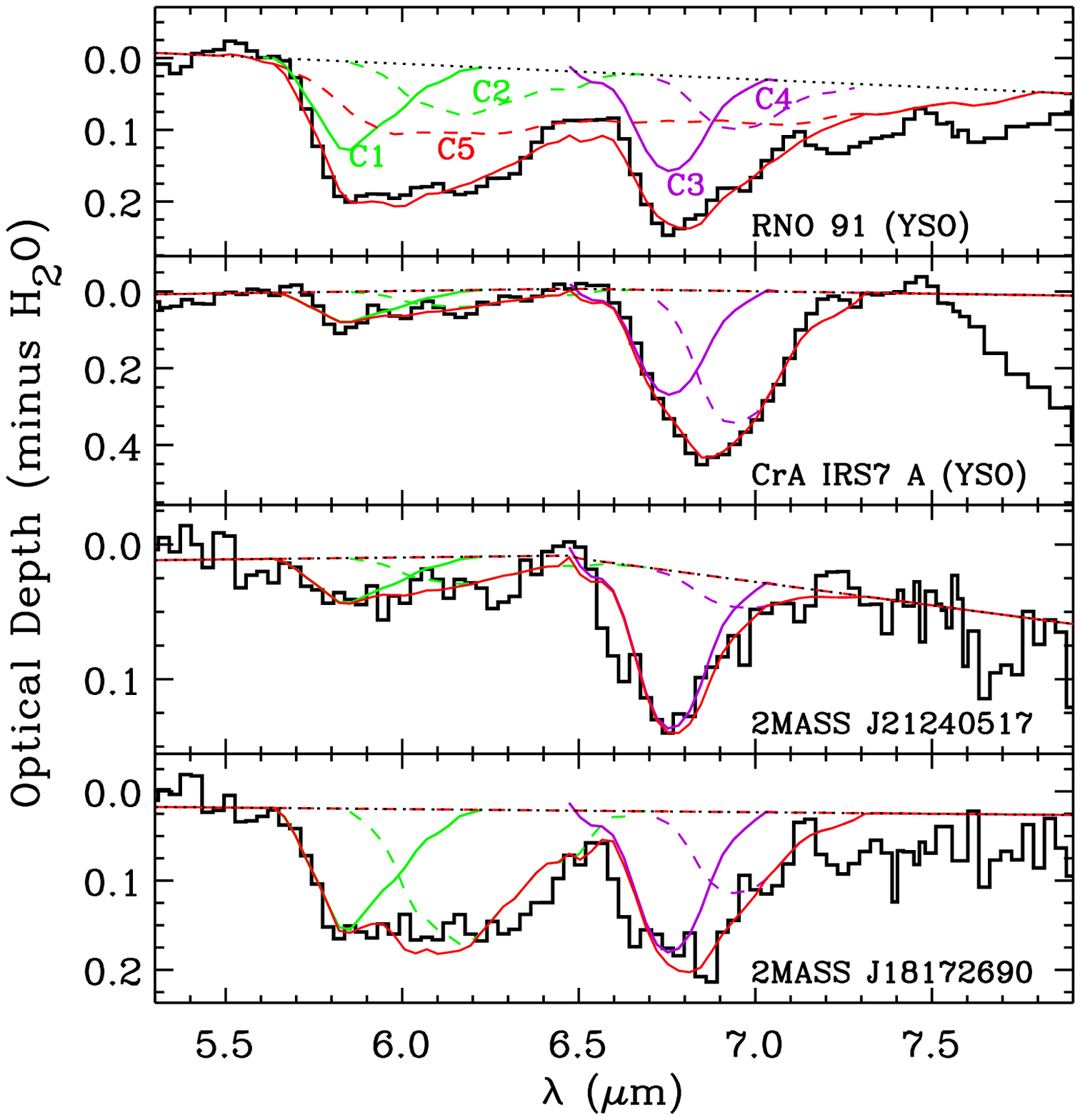} 
  \includegraphics[width=2.55in]{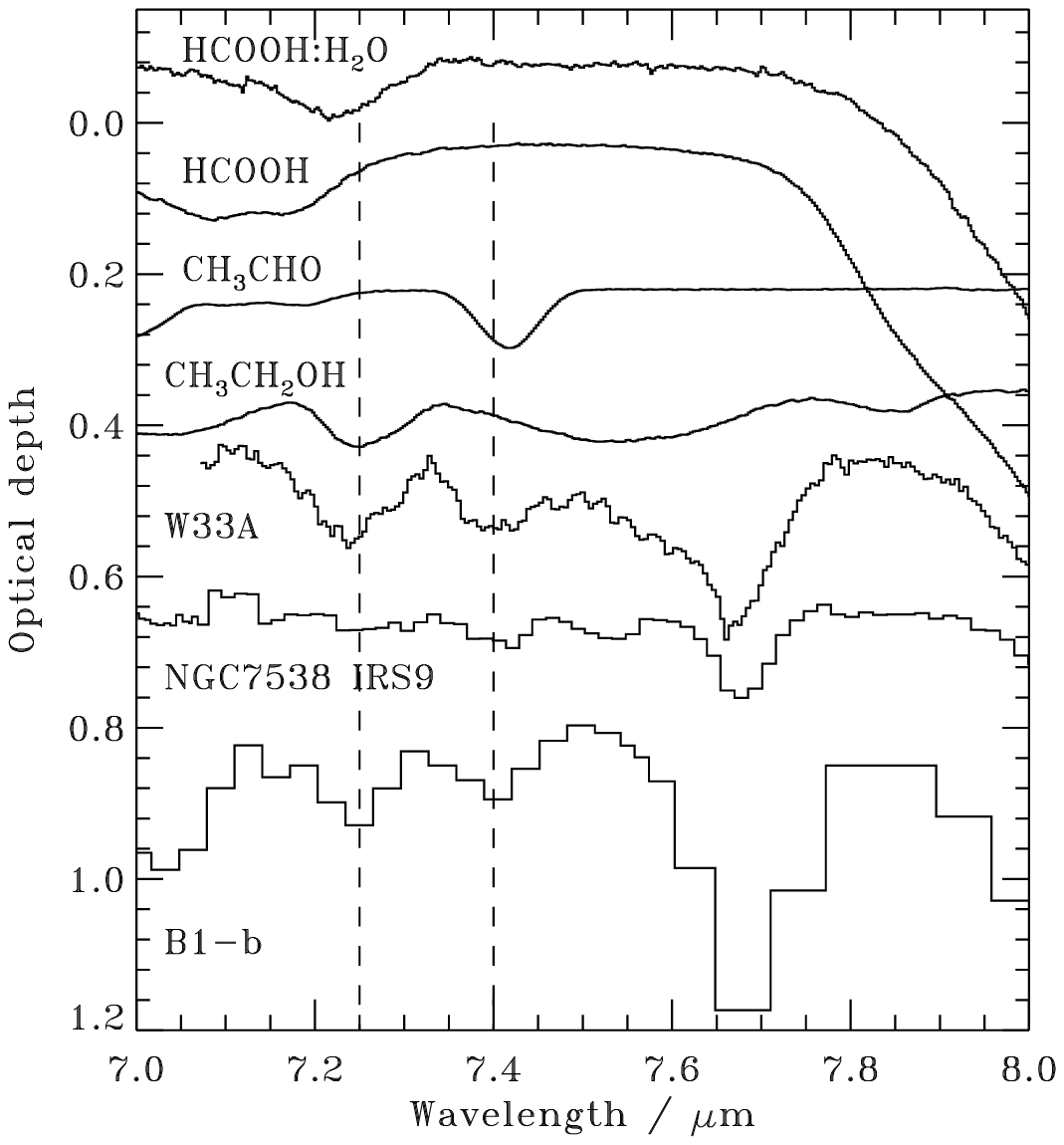} 
 \caption{{\it Left panel:} Decomposition of the absorption features in the 5-8 $\mu$m spectra of two background stars (bottom panels), compared to two YSOs. The sum of the components is indicated with a thick red line. Figure from \cite{Boogert11}. {\it Right panel:} Spectra toward W33A, NGC7538 IRS9 and B1-b, following subtraction of a local spline continuum, plotted together with laboratory spectra of HCOOH, CH$_3$CHO and CH$_3$CH$_2$OH ices. The dashed lines mark the 7.25 and 7.40 $\mu$m features usually assigned to HCOOH and HCOO$^-$.  }
   \label{fig4}
\end{center}
\end{figure}

Figure \ref{fig1} also shows tentative identifications to HCOOH and NH$_4^+$ in a complex band between
5 and 7 $\mu$m. To constrain the carriers of this band  \cite{Boogert08} decomposed it into five different components (C1--C5) after subtraction of the contribution from H$_2$O ice (Fig. \ref{fig4}). From comparison with laboratory ice spectra, C1 has been identified with HCOOH and H$_2$CO, C2 with HCOO$^-$ and NH$_3$, C3 with NH$_4^+$ and CH$_3$OH, C4 with NH$_4^+$ and C5 with warm H$_2$O and anions. The 7.25 $\mu$m has been used to derive HCOOH ice abundances by e.g. \cite{Schutte99}, \cite{Gibb04} and \cite{Boogert08}. The right panel of Figure \ref{fig4} shows that CH$_3$CH$_2$OH is an equally plausible carrier, however. The presence of a 7.25 $\mu$m feature is evidence for the formation of complex ices, but derivations of HCOOH abundances from this band is not possible without higher resolution spectra that can distinguish between the two carriers. 

\section{Ice abundance studies}

Following ice identification, ice column densities are determined by dividing the integrated optical depth of the ice feature by the band strength. Ice abundances are expressed with respect to H$_2$O ice, because H$_2$O ice is predicted to form early during star formation, is the most abundant ice, and has a high sublimation point. Determining ice abundances is important to predict the chemical evolution in different environments and to constrain the C, O and N budget. In addition to these direct uses, the following sections demonstrate how ice abundance statistics can be used to put constraints on how ices form.   Comparison of ice distributions between clouds and protostars constrain which ices require protostellar heating or UV processing to form. Ice abundance variations constrain which ices co-form with H$_2$O and which ices form in a separate stage. Ice correlations provide further constraints on co-formation, and anti-correlations can be used to demonstrate competitive formation. In these investigations it is key to quantify how significant differences and correlations really are. Finally, ice maps provide direct evidence of how ice formation depends on the position of a source in a cloud or a core. 

\subsection{Typical abundances}

\begin{table}
  \begin{center}
  \caption{Recommended ice abundances toward low- and high-mass protostars and clouds.  \label{tab:median}}
  \begin{tabular}{lccc}
  \hline
Ice feature & Low-mass  & High-mass   & background\\
   
   \hline
      H$_2$O 		&100 	&100	&100\\
        CO 			&29		&13		&31\\
       CO$_2$ 		&29		&13		&38\\
     CH$_3$OH 		&3  		&4		&4\\
       NH$_3$ 		& 5 		&5		&-- \\
       CH$_4$ 		& 5		&2		&-- \\
       \vspace{0.2cm}
       XCN 			& 0.3 	&0.6		&-- \\
      pure CO 		&21&3	&--\\
    CO:H$_2$O 		&13&10	&--\\
         CO:CO$_2$ 		& 2	&0.3	&--\\
       pure CO$_2$ 		& 2	&2&--\\
        CO$_2$:H$_2$O 	&20&9&24\\
         CO$_2$:CO 		& 5&5&6\\
       \vspace{0.2cm}
      CO$_2$ shoulder 	& 0.8&1&--\\
            OCN$^-$ 		& 0.2 	&0.6	&--\\
       \vspace{0.2cm}
           2175 cm$^{-1}$	& 0.2 	&0.1&--\\
             C1 (HCOOH + H$_2$CO)$^{\rm a}$ 	& 2.5&2.1&2.8\\
             C2 (HCOO$^{-}$+NH$_3$)$^{\rm a}$ 	& 1.1  &1.3&2.5\\
             C3 (NH$_4^+$ + CH$_3$OH)$^{\rm a}$ 	& 4.3	&4.3	&3.7\\
             C4 (NH$_4^+$)$^{\rm a}$ 	& 2.3 	&4.3&2.1\\
             C5 (warm H$_2$O + anions)$^{\rm a}$ 	&0.9 	&1.4		&--	\\

 \hline
  \end{tabular}
\\$^{\rm a}$Because of multiple band carriers, the reported number is $\tau_{\rm peak} / (N_{\rm H_2O}\times 10^{-18})$.
 \end{center}
\end{table}

Table 1 lists the representative ice abundances found toward low-mass protostars from the $c2d$ survey, high-mass protostars from \cite{Gibb04}and background stars from \cite{Knez05} and \cite{Boogert11}. The reported median abundances are calculated using the Kaplan-Meier (KM) estimate of the survival function. This is a non-parametric procedure that takes into account the constraints provided by upper limits. When calculating the KM estimate, the detections and upper limits are ordered from low to high and the upper limits are given the values of the nearest lower detections. For example in a sample of four detections of 0.5, 1, 3 and 4 and an upper limit of $<$2, the upper limit is treated as a detection of 1. The KM estimate and how to apply statistical tests on it is reviewed by \cite{Feigelson85}. Compared to medians based on only detected ice abundances, the medians calculated from the KM estimate provide more accurate descriptions of ice populations with significant upper limits. The difference is significant,  e.g. the CH$_3$OH and NH$_3$ medians toward high-mass protostars are 40-70\% lower when taking into account the upper limits, resulting in similar abundances toward low- and high-mass YSOs and clouds of 3--5\%. The CO and CO$_2$ abundances vary between the different environments, but are always second in abundance only to H$_2$O.

\subsection{The C, O and N budget}

The amount of C, O and N that are typically bound up in ice mantles is important for the  life cycle of the elements in the interstellar medium.  \cite{Przybilla08} found that the total C, O and N abundances in the solar neighborhood are 2.1, 5.8 and 0.58$\times10^{-4}$ per hydrogen nuclei, respectively. The fractional C, O and N abundance in ices with respect to these total C, O and N abundances can be calculated from the presented ice abundances with respect to H$_2$O ice together with a median H$_2$O abundance with respect to hydrogen of $5\times10^{-5}n_{\rm H}$ from \cite{Pontoppidan04} and \cite{Boogert04}. The resulting median percentages of C, O and N in ices are 8--16\% with respect to the total elemental abundances. When the C and O bound up in grains is subtracted, these values increase to 14--34\% (Table \ref{tab:con}). For the most ice-rich sources in the sample, the amount of O and C in ice approaches the abundances of volatile O and C, indicative of that ice formation only stops when the gas-phase is completely depleted of these elements.

\begin{table}
  \begin{center}
  \caption{The percentages of C, O and N bound up in protostellar ices. \label{tab:con}}
  \begin{tabular}{lcccccc}
\hline
&C$_{\rm ice}$ /  C$_{\rm total}$ &O$_{\rm ice}$ /  O$_{\rm total}$ &N$_{\rm ice}$ /  N$_{\rm total}$ &&C$_{\rm ice}$ /  C$_{\rm vol}$ &O$_{\rm ice}$ /  O$_{\rm vol}$\\
\hline
Low-mass median	&15\%	&16\%	&10\%	&&27\%	&34\%\\
Low-mass max		&46\%	&29\%	&35\%	&&83\%	&61\%\\
High-mass median	&8\%	&12\%	&12\%	&&14\%	&25\%\\
High-mass max	&18\%	&17\%	&22\%	&&32\%	&36\%\\

 \hline
  \end{tabular}
 \end{center}
\end{table}

\begin{figure}[htb]
\begin{center}
 \includegraphics[width=3in]{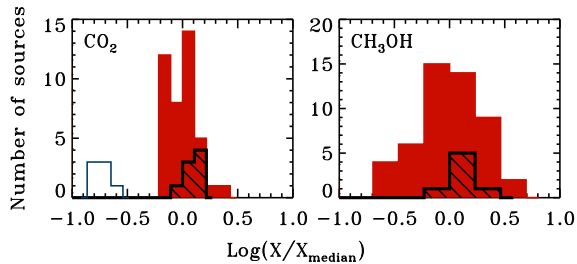} 
 \caption{Histograms of CO$_2$ and CH$_3$OH toward low-mass protostars (red fill), our background star sample (black line-fill) and Taurus background stars (blue contours).}
   \label{fig5}
\end{center}
\end{figure}

\subsection{Cloud versus protostellar ices}

\noindent Comparisons between protostars and background stars provide direct limits on which ices form before the protostar turns on and starts to heat and irradiate its surroundings. \cite{Cook11} showed that in Taurus the CO$_2$ ice fraction in clouds is lower compared to low-mass protostellar envelopes, and proposed that at least some of the CO$_2$ observed toward protostars forms from energetic processing. Similarly, CH$_3$OH ice has been proposed to be a product of protostellar UV ice-processing by e.g. \cite{Gibb04}. Figure \ref{fig5} shows that the CO$_2$ and CH$_3$OH ice abundances toward protostars and background stars from \cite{Boogert11} are similar, and that the Taurus CO$_2$ ice abundances from \cite{Whittet07} are poor templates for ice abundances toward average cloud cores.

\cite{Boogert11} show that there is also no statistically significant difference in the C1-4 components between the protostars and background stars, but the C5 component is absent in cloud sources.The only other feature that is present exclusively toward protostars is pure CO$_2$ ice, which is proposed to form from thermally induced ice segregation or distillation. The protostar thus seems to have a very limited impact on the observable ice composition. This  entails that protostellar ice abundances mainly reflect the ice formation process in the previous cold cloud phase.


\subsection{Protostellar ice abundance variations} 

\begin{figure}[b]
\begin{center}
 \includegraphics[width=2.6in]{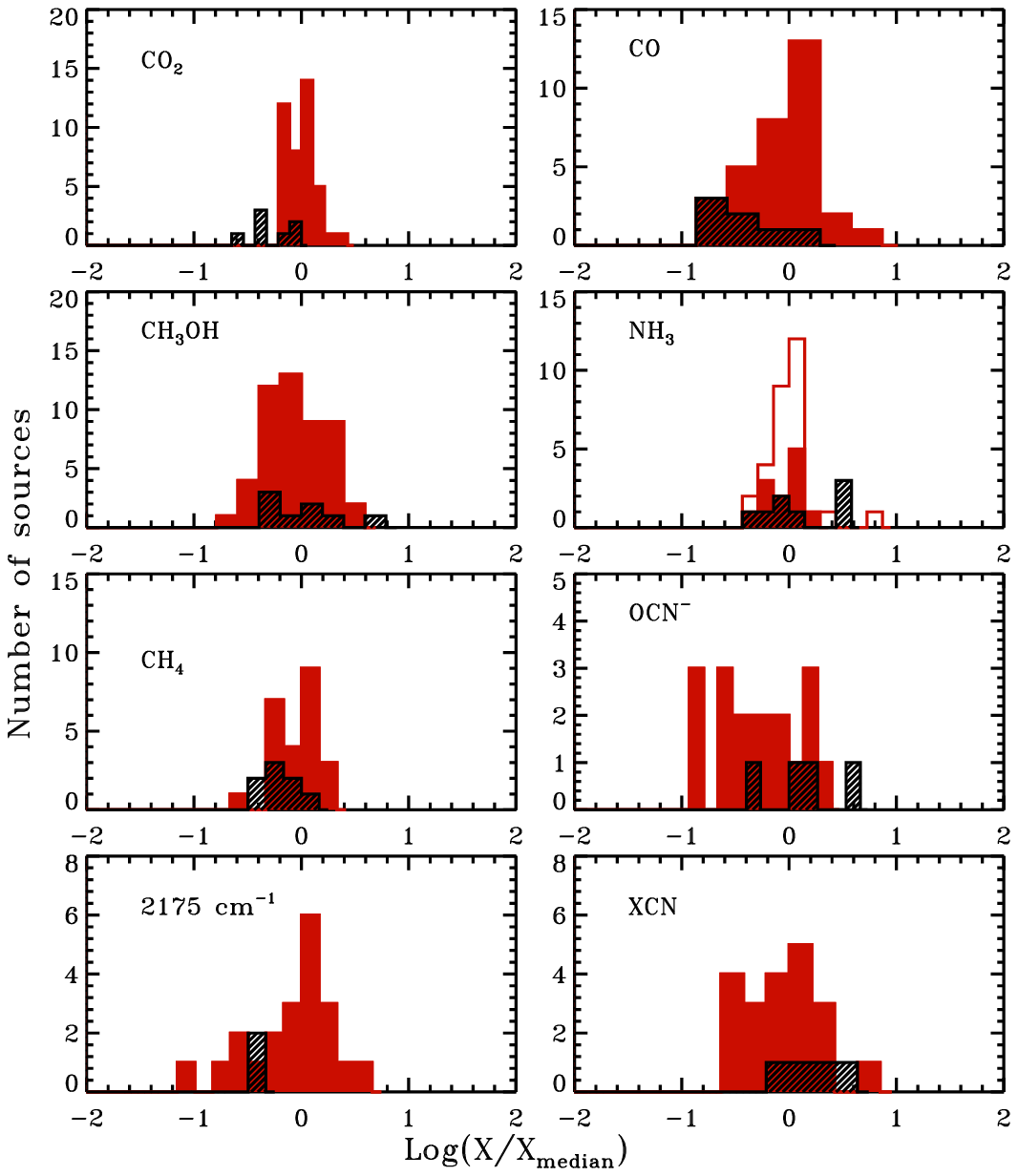} 
 \includegraphics[width=2.6in]{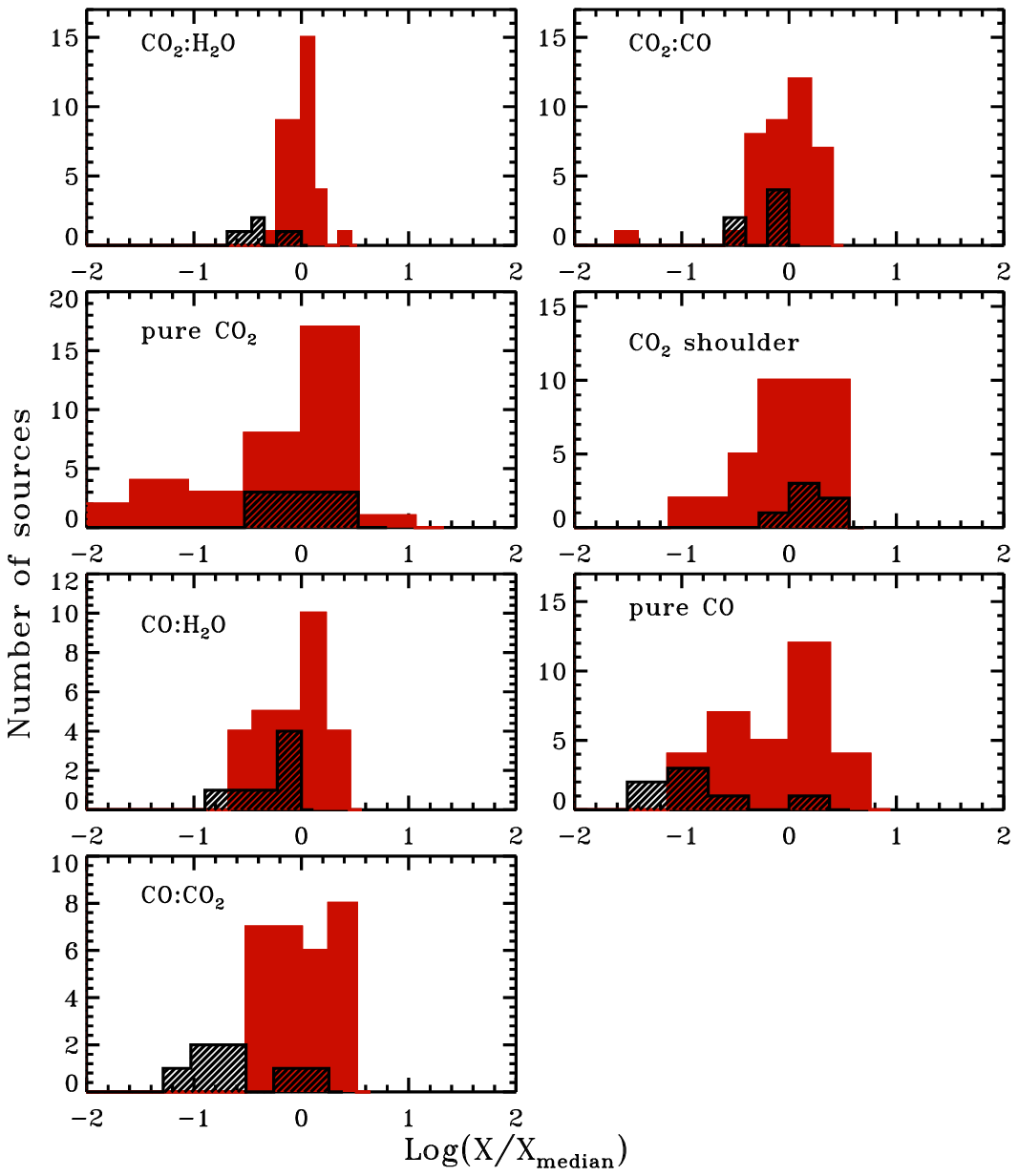} 
 \caption{Histograms of ice and ice component abundances toward low-mass (red) and high-mass (black) protostars. The abundances are log-transformed and normalized to the median detected low-mass protostellar ice abundance. Note the difference in spread between species.}
   \label{fig6}
\end{center}
\end{figure}

Ice abundance variations between different sources depend on how sensitive the ice formation and destruction pathways are to the local environment. Since ice abundances are with respect to H$_2$O, a small variation of a species indicates co-formation with H$_2$O, while large ice abundance variations are indicative of different formation and/or destruction dependencies than H$_2$O ice. Figure \ref{fig6} shows that the CO$_2$, CH$_4$ and NH$_3$ abundance distributions are narrow, indicative of that these ices form together with H$_2$O, while the CO, OCN$^-$ and CH$_3$OH distributions are broad. The ice abundances around high-mass protostars are generally lower compared to low-mass protostars, indicating that heat and UV mainly destroy simple ices. The large CH$_3$OH, CO and OCN$^-$ abundance variations must then originate from a separate prestellar ice formation stage compared to the stage associated with H$_2$O formation. This is consistent with observations that most prestellar and protostellar CO$_2$ is present in a H$_2$O rich ice, while CH$_3$OH and CO ices are not. Figure \ref{fig6} shows the distributions of the individual CO and CO$_2$ components. The pure CO and CO$_2$ ice components have the broadest distributions, consistent with their expected dependence on the protostellar envelope temperature for ice evaporation and segregation. All other CO and CO$_2$ component distributions, except for CO$_2$:H$_2$O, are broad as well, indicative of a different history compared to the H$_2$O-rich ice phase.  

\subsection{Ice maps}

\begin{figure}[htp]
\begin{center}
 \includegraphics[width=3.25in]{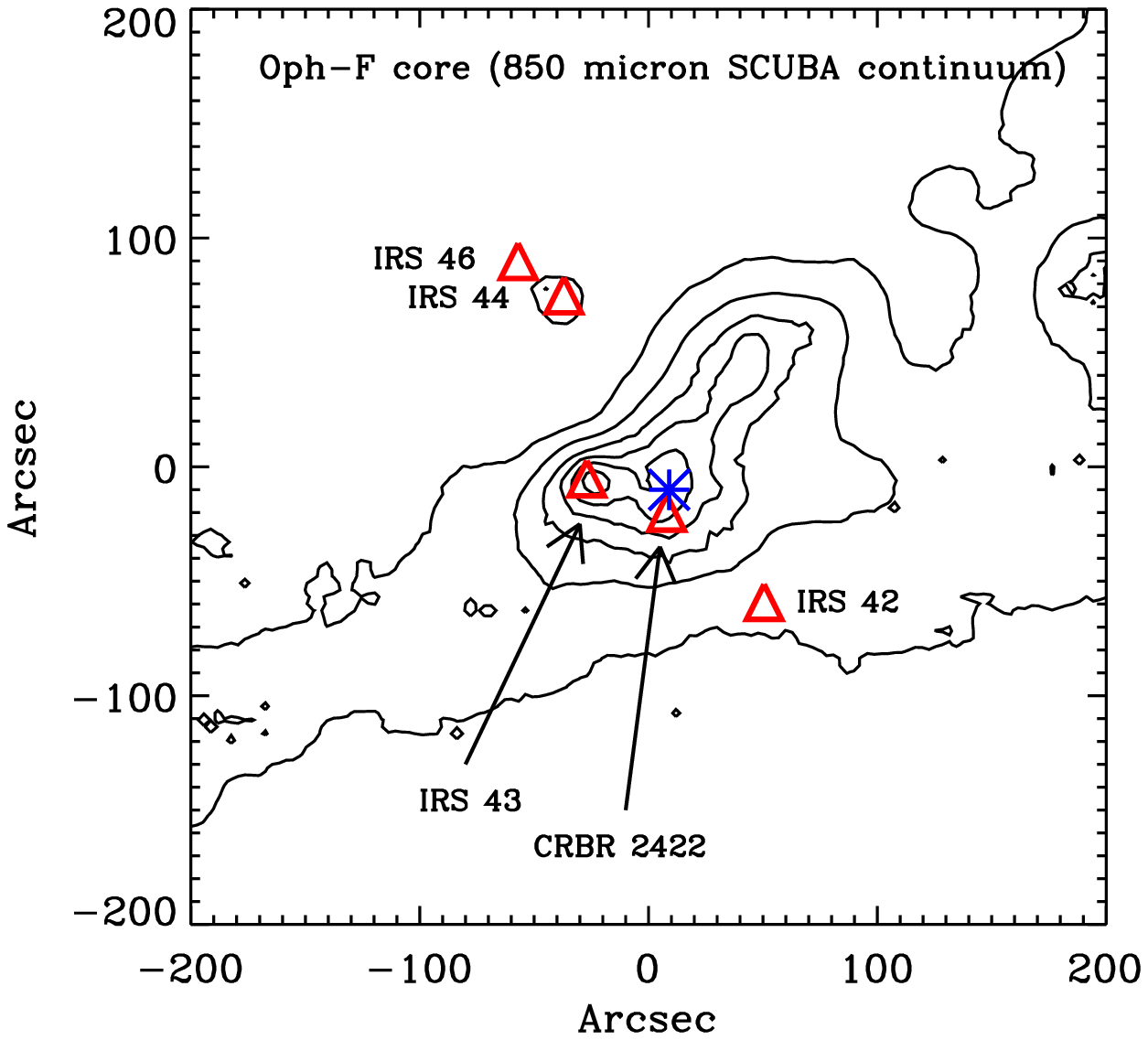}
 \hspace{-0.8in}
 \includegraphics[width=2.75in]{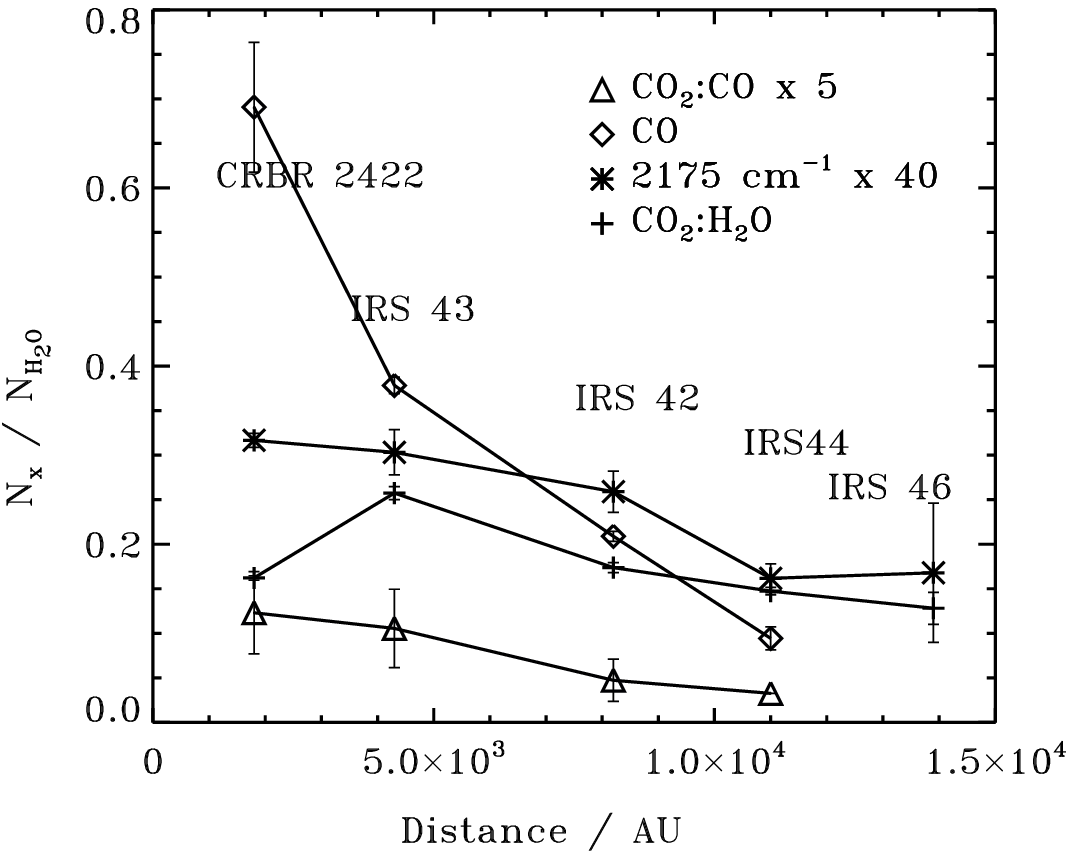}
\caption{{\it Left panel:} The locations of the sources used to construct the ice map plotted on the SCUBA 850 $\mu$m map from \cite{Pontoppidan06} with respect to the central core (star). {\it Right panel:} Ice abundances at different distances from the Oph-F core with respect to H$_2$O ice. The 2175 cm$^{-1}$ abundances are scaled by 40 and the CO$_2$:CO abundances by 5 for clarity.  \label{fig7}}
\end{center}
\end{figure}

The spatial extent of ices can be mapped in clouds and cloud cores if enough background sources are available. Mapping of ice abundances in dense clouds by  \cite{Bergin05} has revealed a similar formation threshold of H$_2$O and CO$_2$ at A$_{\rm V}$=4 mag. CO is only observed at higher extinctions. The H$_2$O-rich ice thus forms first.

\cite{Pontoppidan08} shows that toward the Oph-F core, where protostars are used as background sources to probe the cloud core ices, CO ice abundances increase dramatically toward the core (Fig. \ref{fig7}). \cite{Pontoppidan06} interpreted this as catastrophic freeze-out of CO in the pre-stellar stage once a certain density and temperature is reached. Figure \ref{fig7} shows that the order of magnitude increase in CO ice with respect to H$_2$O toward the core is accompanied by an increase in CO$_2$:CO. In contrast, the CO$_2$:H$_2$O ice component is almost constant. CO:H$_2$O and XCN band are the only other species that increase monotonically toward the densest part of Oph-F core. These bands thus appear directly related to CO freeze-out, indicating that their broad abundance distributions in Fig. 6 are due to their dependence on the prestellar CO freeze-out level. No trend with distance to the core center is seen or expected  for ices that form with H$_2$O (e.g. CH$_4$ and NH$_3$), or of species dependent on protostellar heating (e.g. pure CO$_2$ ice) or of components with potentially multiple carriers, such as the C1-5 bands. CH$_3$OH is only detected toward one of the sources and no trend can be extracted. 

\subsection{Ice abundance correlations}

\begin{figure}[htp]
\begin{center}
 \includegraphics[width=4in]{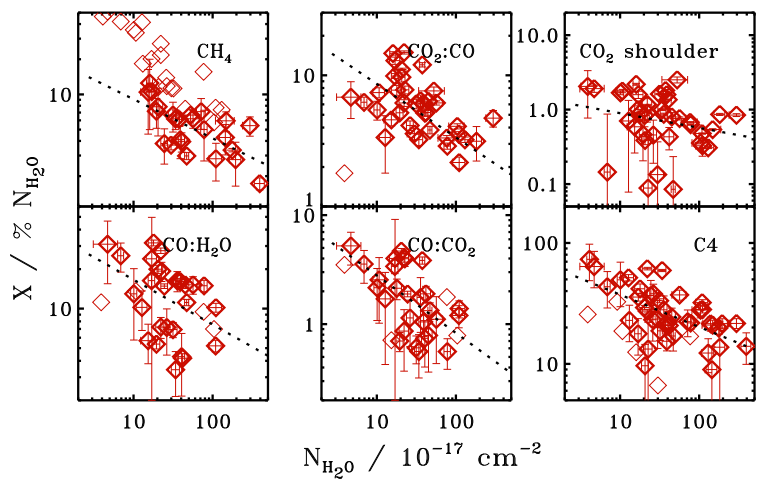}
\caption{Statistically significant correlations between ice abundances w. r. t. H$_2$O (N$_{\rm X}$ /  \% N$_{\rm H_2O}$) and the H$_2$O ice column density for low-mass protostars, with  upper limits plotted with thin symbols. The significance of the correlations was measured using Spearman's rank correlation test, which makes no assumptions about the type of correlation, while the dotted line shows the best log-log fit to the data to guide the eye.\label{fig8}}
\end{center}
\end{figure}

Ice maps are extremely useful, but only possible in a few lines of sight with fortuitous placement of background sources. Ice correlation tests is another tool used to constrain the ice formation history, used by e.g.  \cite{Boogert08} to investigate which ice components depend on ice heating to form. Figure \ref{fig8} shows an example of such a correlation study using the $c2d$ low-mass protostellar sample. Among the investigated ices, the abundance w.r.t H$_2$O of CH$_4$, CO$_2$:CO, the CO$_2$ shoulder, CO:H$_2$O, CO:CO$_2$ and C4 are anti-correlated with the H$_2$O column density at 97--99\% confidence with the rank correlation test introduced by \cite{Spearman04}. The CO-related anti-correlations are consistent with the \S4.5 results that H$_2$O-rich and a CO-rich ice phases form separately in clouds and cloud cores. The anti-correlations are expected if there is competition between the formation of these two phases; i.e. the more oxygen that is bound up in H$_2$O ice, the less may be available to form CO and thus ices that depend on CO freeze-out. Still in each individual core CO ice abundance and H$_2$O ice column are expected to correlate because of increasing CO freeze-out  toward the densest part of the cloud core. This accounts for some of the scatter in the plots. The CH$_4$--H$_2$O trend is strongest for low H$_2$O column densities and may be due to more efficient CH$_4$ formation during the earliest H$_2$O formation stage when then C/CO ratio is high. 

\section{An ice evolution scenario}

Building on the presented results above as well as additional experiments, models and observations, this section discusses what the current constraints on the ice evolution from clouds to protostars.

\begin{figure}[b]
\begin{center}
 \includegraphics[width=4in]{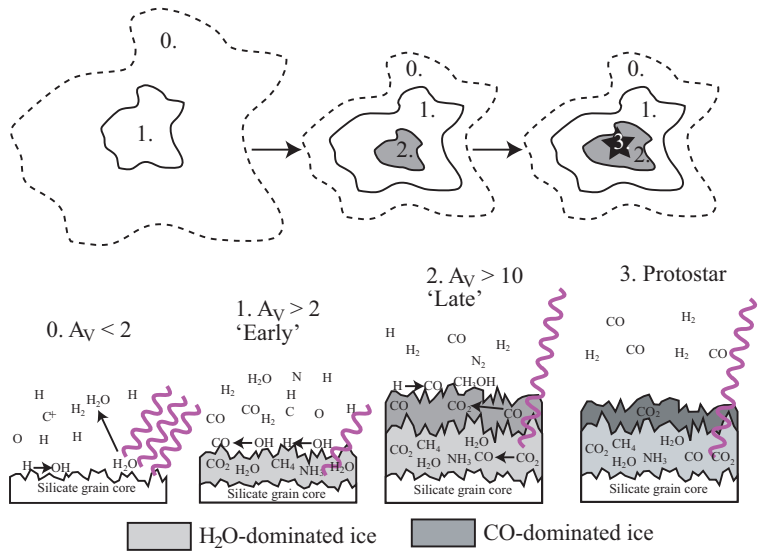}
\caption{The key ice reactions during cloud and star formation, where 0. corresponds to cloud edges where UV photons can penetrate, 1. the early stages of dense cloud formation, 2. the later formation of cloud cores and 3. the protostellar envelope. Ice formation begins in 1. with hydrogenation of atoms resulting in a H$_2$O dominated ice with a high CO$_2$ concentration (CO$_2$:H$_2$O). At later times CO freezes out catastrophically, resulting in a second layer where CO$_2$ formation continues (CO$_2$:CO) and CH$_3$OH formation begins. Small amounts of ice surface chemistry products are maintained in gas-phase due to non-thermal desorption. In the protostellar stage desorption of CO starts at 20~K and H$_2$O:CO$_2$ segregation at 30~K.  \label{fig11}}
\end{center}
\end{figure}

\subsection{Ice formation in clouds and cloud cores}

From ice maps of cloud edges, the first ices to form are H$_2$O dominated with large concentrations of CO$_2$ (Fig. 9). CH$_4$ formation is the most efficient at the earliest stages of H$_2$O ice formation, when the C/CO ratio in the gas phase is still high (\S4.6). The narrow abundance distribution of NH$_3$ is evidence for that it too forms with H$_2$O. This ice formation stage depend on hydrogenation of atoms, O+H, C+H and N+H; CO$_2$ can form efficiently from CO+OH at low temperatures, based on experiments by \cite{Oba10}, \cite{Ioppolo11} and \cite{Noble11}. Most CO that is frozen out in this phase must be converted into CO$_2$ since CO$_2$:H$_2$O is more abundant than CO:H$_2$O. From experiments by \cite{Hiraoka98}, H preferentially reacts with atoms compared to CO, explaining why no CH$_3$OH forms in this ice formation phase.


At some point the CO/O ratio becomes large enough that most frozen-out CO is no longer converted into CO$_2$ and H$_2$O is no longer the most efficiently formed ice. We mark this as the breaking point between early H$_2$O dominated and late CO-driven pre-stellar ice formation. The ices that form in this later stage are not expected to co-vary with H$_2$O, but rather depend on the CO-freeze-out efficiency. The latter is both temperature and density dependent and may vary greatly dependent on cloud structure, cloud collapse time scales and local radiation environment. This explains why many identified ice and ice components that are definitely present during the pre-stellar stage have abundances that vary by orders of magnitude with respect to H$_2$O. In addition, the anti-correlation between H$_2$O ice and the abundance of several CO-ice components suggest that the time available for this CO-based chemistry decreases when a longer time is spent in the H$_2$O-ice formation stage, where O is converted into H$_2$O and CO$_2$:H$_2$O ice .

CO$_2$:CO, CO:H$_2$O and the XCN band all correlate with CO freeze-out in the Oph-F ice map. CO$_2$ may still form from OH+CO at this stage, but a high CO/O ratio results in sparse H$_2$O formation compared to CO and thus in a CO-rich CO$_2$ ice. This is also the outcome from recent modeling by \cite{Garrod11}.  OCN$^{-}$ may form from CO+NH, followed by proton transfer, and the XCN-CO freeze-out correlation supports that OCN$^-$ is responsible for the entire XCN band.   \cite{Cuppen09} have used recent experiments to show that CH$_3$OH and H$_2$CO can form from hydrogenation of CO under ISM conditions and CH$_3$OH abundance should also depend on CO freeze-out. In addition to atomic processing, the ices are continuously exposed to cosmic rays and cosmic-ray induced UV photons --  \cite{Shen04} have estimated the relative doses. This may both affect the simple ice abundances and result in some cold formation of more complex molecules. 

\subsection{Thermal and UV processing of ices around protostars}

\noindent Though much of the observed ice evolution can be explained by processes during the prestellar stage, some ice features are formed or destroyed by the protostar. Based on the observations of pure CO$_2$ by \cite{Pontoppidan08} and experiments by e.g. \cite{Oberg09e}, evaporation of CO from CO$_2$:CO ice and segregation of H$_2$O:CO$_2$ ices are efficient above 20--30~K in protostellar envelopes. Pure CO$_2$ is also observed by  \cite{Kim11} around extremely low-luminosity protostars, however, and may there be a remnant of past outbursts that heated up the envelope beyond its low-accretion temperature, since ice segregation is an irreversible process.  \cite{Evans09} and \cite{Dunham10} have invoked episodic accretion  to explain the luminosity problem of low-mass YSOs, i.e. the observation that most YSOs are significantly less luminous than predicted by protostellar evolutionary models with steady accretion. 

The observed lower CO and CH$_4$ ice abundances toward high-mass protostars compared to low-mass YSOs can also be explained by protostellar ice heating since these are very volatile ices. From experiments by e.g. \cite{Sandford90}, \cite{Collings04} and \cite{Fayolle11}, some CO and CH$_4$ can be trapped in H$_2$O ices, however, which is important for predicting ice abundances in moderately heated regions, such as the comet forming zones in protoplanetary disks. 

From experiments by e.g. \cite{Oberg09d}, moderate ice heating to 20--50~K also enables the movement of radicals within the ice. Radicals are produced in ices by photodissociation of the condensed molecules. Models by \cite{Garrod08} predict that especially CH$_3$OH photodissociation followed by radical diffusion result in complex ice formation  and protostellar envelopes are thus expected to be efficient factories of complex organics. UV radiation of CO:CH$_3$OH ice mixtures produce a more HCO-rich complex chemistry, which may dominate at lower temperatures when CO is still frozen out. The ice feature at 7.25~$\mu$m identified with to HCOOH or CH$_3$CH$_2$OH provides direct evidence for this complex ice chemistry. Most complex ice features cannot be directly observed, however, because of band confusion. Instead most constraints on this ice evolutionary phase comes from millimeter observations of desorbed ice chemistry products.

Complex organic molecules such as CH$_3$CH$_2$OH and HCOOCH$_3$ are observed in the inner regions around high-mass protostars, in so called hot cores, at abundances that cannot be reproduced by gas phase chemistry. Furthermore, \cite{Bisschop07} found that they correlate with the ice chemistry product CH$_3$OH. These observational pieces of evidence, together with the existence of a plausible ice production pathway of complex molecules, strongly support a scenario in which complex ices are produced in the outer envelope and then desorb into the gas-phase as icy grains flow in towards the protostar. 

IRAS 16293-2422,  observed by e.g. \cite{vanDishoeck95} and \cite{ Cazaux03} is a low-mass equivalent to the high-mass hot cores, and there are at least five other low-mass protostars where complex molecules such as HCOOCH$_3$ and CH$_3$OCH$_3$ are detected by \cite{Bottinelli04b}, \cite{Jorgensen05} and \cite{vanderMarel11}. \cite{Arce08} also detected complex molecules in a low-mass outflow and \cite{Oberg10b} recently detected HCOOCH$_3$ in  a cold and quiescent part of a low-mass star forming region that probably has been illuminated by UV through a protostellar outflow cavity. Combining these observational results, \"Oberg et al. (2011b, subm. to ApJ) find that HCOOCH$_3$ generally dominates the complex chemistry toward outflows and low-mass protostellar envelopes, consistent with experiments on CO:CH$_3$OH ice photochemistry below 25~K. In contrast, observations of   high-mass and resolved low-mass hot cores contain more CH$_3$OCH$_3$ or C$_2$H$_5$OH than HCOOCH$_3$. This is evidence for a sequential formation of complex molecules, starting with HCO-rich molecules as long as CO ice is abundant, followed by CH$_{3/2}$-rich molecules at higher ice temperatures. Gas phase observations can thus provide information on the ice chemistry, but interpretations of these more indirect constraints rely on a combination of experimental and model results.

\section{Conclusions}

Thanks to a combination of ground-based and space telescope ice observations and years of laboratory ice spectroscopy and ice chemistry experiments we have good constraints on the typical ice abundances and ice compositions in a range of interstellar and circumstellar environments. A key result coming out of these observations is that most ices up to CH$_3$OH in complexity have already formed by the time the protostar turns on. Ice abundance variations must therefore be mainly due to variations in the prestellar ice formation processes. Variations in CO freeze-out and the subsequent CO-based chemistry is a likely main contributor to the observed differences in ice abundances in different lines of sight, but other processes may be important as well. Thermal and UV processing of ices by the protostar is important for the continued evolution of the ice, leading up to the formation of hot core molecules and the complex organics found in comets. To further constrain the ice evolution from cloud cores to comets will require a combination of high-resolution IR studies to detect complex molecules directly in the ice, millimeter observations toward protostars and disks to image desorbed ice chemistry products, and more detailed experiments and models of both the early ice formation stages and of the reactions that convert simple ices into more complex ones.


\begin{thebibliography}{63}
\expandafter\ifx\csname natexlab\endcsname\relax\def\natexlab#1{#1}\fi

\bibitem[{{Arce} {et~al.}(2008)}]{Arce08}
{Arce}, H.~G., {Santiago-Garc{\'{\i}}a}, J., {J{\o}rgensen}, et al. 2008, ApJl, 681, L21
  
\bibitem[{{Bergin} {et~al.}(2005)}]{Bergin05}
{Bergin}, E.~A., {Melnick}, G.~J., {Gerakines}, P.~A., et al. 2005, ApJl, 627, L33


\bibitem[{{Bisschop} {et~al.}(2007)}]{Bisschop07}
{Bisschop}, S.~E., {J{\o}rgensen}, J.~K., {van Dishoeck}, E.~F. et al. 2007{\natexlab{b}}, A\&A, 465, 913


\bibitem[{{Boogert} \& {Ehrenfreund}(2004)}]{Boogert04}
{Boogert}, A.~C.~A. \& {Ehrenfreund}, P. 2004, in ASP Conf. Ser. 309:
  Astrophysics of Dust,  547--572

\bibitem[{{Boogert} {et~al.}(2011)}]{Boogert11}
{Boogert}, A.~C.~A., {Huard}, T.~L., {Cook}, A.~M., {et~al.} 2011, ApJ, 729, 92

\bibitem[{{Boogert} {et~al.}(2008)}]{Boogert08}
{Boogert}, A.~C.~A., {Pontoppidan}, K.~M., {Knez}, C., {et~al.} 2008, ApJ, 678,
  985

\bibitem[{{Boogert} {et~al.}(2004)}]{Boogert04b}
{Boogert}, A.~C.~A., {Pontoppidan}, K.~M., {Lahuis}, F., {et~al.} 2004, ApJs,
  154, 359

\bibitem[{{Bottinelli} {et~al.}(2010)}]{Bottinelli10}
{Bottinelli}, S., {Adwin Boogert}, A.~C., {Bouwman}, J., {et~al.} 2010, ApJ,
  718, 1100

\bibitem[{{Bottinelli} {et~al.}(2004)}]{Bottinelli04b}
{Bottinelli}, S., {Ceccarelli}, C., {Neri}, R., {et~al.} 2004, ApJl, 617, L69

\bibitem[{{Cazaux} {et~al.}(2003)}]{Cazaux03}
{Cazaux}, S., {Tielens}, A.~G.~G.~M., {Ceccarelli}, C., {et~al.} 2003, ApJl,
  593, L51


\bibitem[{{Charnley}(2004)}]{Charnley04}
{Charnley}, S.~B. 2004, Advances in Space Res., 33, 23

\bibitem[{{Charnley} {et~al.}(1992)}]{Charnley92}
{Charnley}, S.~B., {Tielens}, A.~G.~G.~M., \& {Millar}, T.~J. 1992, ApJl, 399,
  L71


\bibitem[{{Collings} {et~al.}(2004)}]{Collings04}
{Collings}, M.~P., {Anderson}, M.~A., {Chen}, R., {et~al.} 2004, MNRAS, 354,
  1133

\bibitem[{{Cook} {et~al.}(2011)}]{Cook11}
{Cook}, A.~M., {Whittet}, D.~C.~B., {Shenoy}, S.~S., {et~al.} 2011, ApJ, 730,
  124

\bibitem[{{Cuppen} {et~al.}(2009)}]{Cuppen09}
{Cuppen}, H.~M., {van Dishoeck}, E.~F., {Herbst}, E., \& {Tielens}, A.~G.~G.~M.
  2009, A\&A, 508, 275


\bibitem[{{D'Hendecourt} \& {Jourdain de Muizon}(1989)}]{dHendecourt89}
{D'Hendecourt}, L.~B. \& {Jourdain de Muizon}, M. 1989, A\&A, 223, L5

\bibitem[{{Dunham} {et~al.}(2010)}]{Dunham10}
{Dunham}, M.~M., {Evans}, N.~J., {Bourke}, T.~L., {et~al.} 2010, ApJ, 721, 995

\bibitem[{{Evans} {et~al.}(2009)}]{Evans09}
{Evans}, N.~J., {Dunham}, M.~M., {J{\o}rgensen}, J.~K., {et~al.} 2009, ApJs,
  181, 321

\bibitem[{{Evans} {et~al.}(2003)}]{Evans03}
{Evans}, II, N.~J., {Allen}, L.~E., {Blake}, G.~A., {et~al.} 2003, PASP, 115,
  965

\bibitem[{{Fayolle} {et~al.}(2011)}]{Fayolle11}
{Fayolle}, E.~C., {{\"O}berg}, K.~I., {Cuppen}, H.~M., {Visser}, R., \&
  {Linnartz}, H. 2011, A\&A, 529, A74+

\bibitem[{{Feigelson} \& {Nelson}(1985)}]{Feigelson85}
{Feigelson}, E.~D. \& {Nelson}, P.~I. 1985, ApJ, 293, 192

\bibitem[{{Fraser} {et~al.}(2005)}]{Fraser05}
{Fraser}, H.~J., {Bisschop}, S.~E., {Pontoppidan}, K.~M., et al. 2005, MNRAS, 356, 1283


\bibitem[{{Garrod} \& {Pauly}(2011)}]{Garrod11}
{Garrod}, R.~T. \& {Pauly}, T. 2011, ArXiv e-prints

\bibitem[{{Garrod} {et~al.}(2008)}]{Garrod08}
{Garrod}, R.~T., {Weaver}, S.~L.~W., \& {Herbst}, E. 2008, ApJ, 682, 283

\bibitem[{{Gibb} {et~al.}(2004)}]{Gibb04}
{Gibb}, E.~L., {Whittet}, D.~C.~B., {Boogert}, A.~C.~A., \& {Tielens},
  A.~G.~G.~M. 2004, ApJs, 151, 35

\bibitem[{{Gibb} {et~al.}(2000)}]{Gibb00}
{Gibb}, E.~L., {Whittet}, D.~C.~B., {Schutte}, W.~A., {et~al.} 2000, ApJ, 536,
  347

\bibitem[{{Gillett} \& {Forrest}(1973)}]{Gillet73}
{Gillett}, F.~C. \& {Forrest}, W.~J. 1973, ApJ, 179, 483

\bibitem[{{Hiraoka} {et~al.}(1998)}]{Hiraoka98}
{Hiraoka}, K., {Miyagoshi}, T., {Takayama}, T., {Yamamoto}, K., \& {Kihara}, Y.
  1998, ApJ, 498, 710


\bibitem[{{Ioppolo} {et~al.}(2011)}]{Ioppolo11}
{Ioppolo}, S., {van Boheemen}, Y., {Cuppen}, H.~M., et al. H. 2011, MNRAS, 238

\bibitem[{{J{\o}rgensen} {et~al.}(2005)}]{Jorgensen05}
{J{\o}rgensen}, J.~K., {Bourke}, T.~L., {Myers}, P.~C., {et~al.} 2005, ApJ,
  632, 973

\bibitem[{{Keane} {et~al.}(2001)}]{Keane01}
{Keane}, J.~V., {Boonman}, A.~M.~S., {Tielens}, A.~G.~G.~M., et al. 2001, A\&A, 376, L5

\bibitem[{{Kim.} {et~al.}(2011)}]{Kim11}
{Kim.}, H.~J., {Evans}, II, N.~J., {Dunham}, M.~M., \& {Lee}, J.~E. 2011, in
  IAU Symposium, 280, 216

\bibitem[{{Knez} {et~al.}(2005)}]{Knez05}
{Knez}, C., {Boogert}, A.~C.~A., {Pontoppidan}, K.~M., {et~al.} 2005, ApJl,
  635, L145

\bibitem[{{Merrill} {et~al.}(1976)}]{Merrill76}
{Merrill}, K.~M., {Russell}, R.~W., \& {Soifer}, B.~T. 1976, ApJ, 207, 763


\bibitem[{{Oba} {et~al.}(2010)}]{Oba10}
{Oba}, Y., {Watanabe}, N., {Kouchi}, A., {Hama}, T., \& {Pirronello}, V. 2010,
  ApJl, 712, L174

\bibitem[{{{\"O}berg} {et~al.}(2008)}]{Oberg08}
{{\"O}berg}, K.~I., {Boogert}, A.~C.~A., {Pontoppidan}, K.~M., {et~al.} 2008,
  ApJ, 678, 1032

\bibitem[{{{\"O}berg} {et~al.}(2010)}]{Oberg10b}
{{\"O}berg}, K.~I., {Bottinelli}, S., {J{\o}rgensen}, J.~K., \& {van Dishoeck},
  E.~F. 2010, ApJ, 716, 825

\bibitem[{{{\"O}berg} {et~al.}(2009b)}]{Oberg09e}
{{\"O}berg}, K.~I., {Fayolle}, E.~C., {Cuppen}, H.~M., et al., H. 2009{\natexlab{a}}, A\&A, 505, 183

\bibitem[{{{\"O}berg} {et~al.}(2009a)}]{Oberg09d}
{{\"O}berg}, K.~I., {Garrod}, R.~T., {van Dishoeck}, E.~F., \& {Linnartz}, H.
  2009, A\&A, 504, 891

\bibitem[{{Noble} {et~al.}(2011)}]{Noble11}
{Noble}, J.~A., {Dulieu}, F., {Congiu}, E., \& {Fraser}, H.~J. 2011, ApJ, 735,
  121

\bibitem[{{Pendleton} {et~al.}(1999)}]{Pendleton99}
{Pendleton}, Y.~J., {Tielens}, A.~G.~G.~M., {Tokunaga}, A.~T., et al. 1999, ApJ, 513, 294

\bibitem[{{Pontoppidan}(2006)}]{Pontoppidan06}
{Pontoppidan}, K.~M. 2006, A\&A, 453, L47

\bibitem[{{Pontoppidan} {et~al.}(2008)}]{Pontoppidan08}
{Pontoppidan}, K.~M., {Boogert}, A.~C.~A., {Fraser}, H.~J., {et~al.} 2008, ApJ,
  678, 1005

\bibitem[{{Pontoppidan} {et~al.}(2003)}]{Pontoppidan03b}
{Pontoppidan}, K.~M., {Dartois}, E., {van Dishoeck}, E.~F., et al. 2003{\natexlab{a}}, A\&A, 404, L17

\bibitem[{{Pontoppidan} {et~al.}(2003)}]{Pontoppidan03}
{Pontoppidan}, K.~M., {Fraser}, H.~J., {Dartois}, E., {et~al.}
  2003{\natexlab{b}}, A\&A, 408, 981

\bibitem[{{Pontoppidan} {et~al.}(2004)}]{Pontoppidan04}
{Pontoppidan}, K.~M., {van Dishoeck}, E.~F., \& {Dartois}, E. 2004, A\&A, 426,
  925

\bibitem[{{Przybilla} {et~al.}(2008)}]{Przybilla08}
{Przybilla}, N., {Nieva}, M., \& {Butler}, K. 2008, ApJl, 688, L103

\bibitem[{{Reach} {et~al.}(2009)}]{Reach09}
{Reach}, W.~T., {Faied}, D., {Rho}, J., {et~al.} 2009, ApJ, 690, 683



\bibitem[{{Sandford} \& {Allamandola}(1990)}]{Sandford90}
{Sandford}, S.~A. \& {Allamandola}, L.~J. 1990, ApJ, 355, 357

\bibitem[{{Schutte} {et~al.}(1999)}]{Schutte99}
{Schutte}, W.~A., {Boogert}, A.~C.~A., {Tielens}, A.~G.~G.~M., {et~al.} 1999,
  A\&A, 343, 966


\bibitem[{{Shen} {et~al.}(2004)}]{Shen04}
{Shen}, C.~J., {Greenberg}, J.~M., {Schutte}, W.~A., \& {van Dishoeck}, E.~F.
  2004, A\&A, 415, 203
  
  

\bibitem[{Spearman(1904)}]{Spearman04}
Spearman, C. 1904, The American Journal of Psychology, 15, pp. 72


\bibitem[{{Tielens} \& {Hagen}(1982)}]{Tielens82}
{Tielens}, A.~G.~G.~M. \& {Hagen}, W. 1982, A\&A, 114, 245

\bibitem[{{Tielens} {et~al.}(1991)}]{Tielens91}
{Tielens}, A.~G.~G.~M., {Tokunaga}, A.~T., {Geballe}, T.~R., \& {Baas}, F.
  1991, ApJ, 381, 181


\bibitem[{{van der Marel} {et~al.}(2011)}]{vanderMarel11}
{van der Marel}, N., {\"Oberg}, K.~I., {Kristensen}, L., et al.,  2011, IAU Symposium 280, 365

\bibitem[{{van Broekhuizen} {et~al.}(2004)}]{vanBroekhuizen04}
{van Broekhuizen}, F.~A., {Keane}, J.~V., \& {Schutte}, W.~A. 2004, A\&A, 415,
  425

\bibitem[{{van Broekhuizen} {et~al.}(2005)}]{vanBroekhuizen05}
{van Broekhuizen}, F.~A., {Pontoppidan}, K.~M., {Fraser}, H.~J., et al. 2005, A\&A, 441, 249

\bibitem[{{van Dishoeck} {et~al.}(1995)}]{vanDishoeck95}
{van Dishoeck}, E.~F., {Blake}, G.~A., {Jansen}, D.~J., \& {Groesbeck}, T.~D.
  1995, ApJ, 447, 760


\bibitem[{{Whittet} {et~al.}(2009)}]{Whittet09}
{Whittet}, D.~C.~B., {Cook}, A.~M., {Chiar}, J.~E., {et~al.} 2009, ApJ, 695, 94

\bibitem[{{Whittet} {et~al.}(2007)}]{Whittet07}
{Whittet}, D.~C.~B., {Shenoy}, S.~S., {Bergin}, E.~A., {et~al.} 2007, ApJ, 655,
  332

\bibitem[{{Zasowski} {et~al.}(2009)}]{Zasowski09}
{Zasowski}, G., {Kemper}, F., {Watson}, D.~M., {et~al.} 2009, ApJ, 694, 459

\end{thebibliography}

\begin{discussion}

\discuss{Amanda Cook}{At least once in your talk you referred to a pure CO$_2$ component in the ice. Since it seems like CO$_2$ and H$_2$O forms in tandem in molecular clouds, how would you explain the presence of pure CO$_2$ ice?} 

\discuss{\"Oberg}{Pure CO$_2$ ice can form from evaporation of CO at 20~K from the CO$_2$:CO ice layer that is a result of late CO$_2$ formation in cloud cores, or from segregation of the H$_2$O:CO$_2$ ice layer starting at 30~K. Pure CO$_2$ is therefore only expected in environments where ices have been heated to at least 20--30~K.}

\discuss{Name Sakai}{What is the origin of the difference of molecular abundances in the ice between low-mass and high-mass sources? Are the ice abundances in high-mass cases measured in pre-stellar cores?}

\discuss{\"Oberg}{The high-mass sources are all protostars. Low CO and CH$_4$ ice abundances toward high-mass YSOs are probably due to that the protostars have heated most of the envelopes, resulting in evaporation of volatile ices. The difference in CO$_2$ ice abundances may be due to either CO$_2$ evaporation in the high-mass protostellar envelope, or due to less CO$_2$ formation in the high-mass pre-stellar phase because of different collapse time scales, and density and temperature structures in low- and high-mass prestellar cores}

\discuss{Jacob Laas}{The recent studies by Gerakines et al. have found that NH$_3$ and other N-bearing species can have large effects on certain line profiles in ices. Could this have any effect on the spectral signatures near 7$\mu$m that you say are strong indicators of complex organic molecules?}

\discuss{\"Oberg}{NH$_3$ acts as a base in ices, transforming e.g. HCOOH into HCOO$^{-}$. If the two bands at 7.2--7.5 $\mu$m are due (partially) to HCOOH and HCOO$^{-}$, NH$_3$ may regulate the relative intensities of the two bands. If the bands are instead due to CH$_3$CH$_2$OH and CH$_3$CHO, NH$_3$ should not have a large effect on the band characteristics.}

\end{discussion}

\end{document}